\documentclass[namedreferences]{solarphysicsnew}
\usepackage[optionalrh]{spr-sola-addons} % For Solar Physics
\usepackage{graphicx}        % For eps figures, newer & more powerfull
\usepackage{color}           % For color text: \color command
\usepackage{url}             % For breaking URLs easily trough lines
\ifx \doiurl
 \undefined \def
\doiurl#1{\href{http://dx.doi.org/#1}{\textsf{DOI}}}
\fi

\begin{document}

\begin{article}

\begin{opening}

\title{Brightness Temperature of Radio Zebras and Wave Energy Densities in Their Sources}

\author{L.V. \surname{Yasnov}$^{1}$\sep
        J. \surname{Ben\'a\v{c}ek}$^{2}$\sep
        M.   \surname{Karlick\'y}$^{3}$}

        \runningauthor{L.V. Yasnov \textit{et al.}}
        \runningtitle{Brightness Temperature of Radio Zebras}

%--------------------------------------------------------------------------------------------------------------------------------------------------------
\institute{$^{1}$ St.-Petersburg State
                  University, St.-Petersburg, 198504, Russia
                  \email{l.yasnov@spbu.ru}\\
           $^{2}$ Department of Theoretical Physics and Astrophysics, Masaryk University,
   Kotl\'a\v{r}sk\'a 2, CZ -- 611 37 Brno, Czech Republic
           \email{jbenacek@physics.muni.cz}\\
           $^{3}$ Astronomical Institute, Academy of Sciences of
           the Czech Republic, 251 65 Ond\v{r}ejov, Czech Republic
           \email{karlicky@asu.cas.cz}\\
             }

\date{Received ; accepted }

%--------------------------------------------------------------------------------------------------------------------------------------------------------
%Abstract
%

\begin{abstract}
We estimated the brightness temperature of radio zebras (zebra pattern -- ZP),
considering that ZPs are generated in loops having an exponential density
profile in their cross-section. We took into account that when in plasma there
is a source emitting in all directions, then in the escape process from the
plasma the emission obtains a directional character nearly perpendicular to the
constant-density profile. Owing to the high directivity of the plasma emission
(for emission at frequencies close to the plasma frequency) the region from
which the emission escapes can be very small. We estimated the brightness
temperature of three observed ZPs for two values of the density scale height (1
and 0.21 Mm) and two values of the loop width (1 and 2 arcsec). In all cases
high brightness temperatures were obtained. For the higher value of the density
scale height, the brightness temperature was estimated as 1.1 $\times$
10$^{15}$ -- 1.3 $\times$ 10$^{17}$~K, and for the lower value as 4.7 $\times$
10$^{13}$ -- 5.6 $\times$ 10$^{15}$~K. These temperatures show that the
observation probability of a burst with ZP, which is generated in the
transition region with a steep gradient of the plasma density, is significantly
higher than for a burst generated in a region with smoother changes of the
plasma density.  We also computed the saturation energy density of the
upper-hybrid waves (which according to the double plasma resonance model are
generated in the zebra source) using a 3D particle-in-cell model with the
loss-cone type of distribution of hot electrons. We found that this saturated
energy is proportional to the ratio of hot electron and background plasma
densities. Thus, comparing the growth rate and collisional damping of the
upper-hybrid waves, we estimated minimal densities of hot electrons as well as
the minimal value of the saturation energy density of the upper-hybrid waves.
Finally, we compared the computed energy density of the upper-hybrid waves with
the energy density of the electromagnetic waves in the zebra source and thus
estimated the efficiency of the wave transformation.
\end{abstract}

%--------------------------------------------------------------------------------------------------------------------------------------------------------
%Keywords
%
\end{opening}

%--------------------------------------------------------------------------------------------------------------------------------------------------------
\section{Introduction}

Fine structures of solar radio bursts are very important for understanding
flare energy-release processes and diagnostics of the flare plasma. Among various fine structures the most intriguing one is the zebra structure (ZP --
zebra pattern) occurring in Type IV radio bursts. In radio spectra it appears as
several parallel stripes distributed uniformly in frequency; see examples
bellow. Usually the number of such zebra stripes in ZP is large ($>
5-8$, sometimes even exceeding $20$).

There are still questions about the generation mechanism of these
ZPs. Among many proposed models (\opencite{1975SoPh...44..461Z};
\opencite{2003ApJ...593.1195L}; \opencite{2006A&A...450..359B};
 \opencite{2007SoPh..241..127K}; \opencite{2010Ap&SS.325..251T};
\opencite{2013A&A...552A..90K}), the most commonly accepted model is that based
on the double-plasma resonance (DPR) (\opencite{2013SoPh..284..579Z};
\opencite{2015A&A...581A.115K}). Based on this model, most of the observed
characteristics of ZPs were explained: the frequency range, polarization,
amount of stripes and their frequencies, their high-frequency limit, and their
temporal changes.

However, up to now in the literature there are only a few estimations of the ZP
brightness temperature, which is important for further specification of the
generation mechanism of ZPs. For example,~\inlinecite{1994SoPh..155..373C}
estimated the brightness temperature of metric ZPs to be 10$^{10}$ K with the
source size constrained by the Nan\c{c}ay Radioheliograph.

Further estimation of the ZP brightness temperature (T$_{\rm b}$ = 10$^{13}$ K)
was by~\inlinecite{2003A&A...406.1071C}, where the ZP consisted of spiky
superfine structures; see also~\inlinecite{2006SSRv..127..195C}. On the other
hand, ~\inlinecite{2006SSRv..127..195C} states that the metric ZP radio sources
occupy a noticeable part of the background continuum source or even the entire
active region. The half-width of one source of the metric ZP was about 1.9
arcmin, which gives a brightness temperature of 10$^{10}$ K.

 Using the \textit{Siberian Solar Radio Telescope}
(SSRT),  \inlinecite{2005A&A...431.1037A} observed a ZP burst at $\approx$ 5.7
GHz (the highest frequency ever reported for ZP emission), which yielded a
lower limit of T$_{\rm b}$ $\approx$ 2 $\times$ 10$^7$ K.

\inlinecite{2011ApJ...736...64C} estimated the lower limit for the
decimetric ZP brightness temperature as 1.6 $\times$ 10$^9$ K.
Finally, ~\inlinecite{2014ApJ...790..151T} estimated the brightness temperature
of a decimetric ZP as T$_{\rm b}$ $\approx$ 2 $\times$ 10$^{11}$~K.

In the present article we estimate the brightness temperature of ZP considering
that a ZP is generated in the loop having in its cross-section an exponential
density profile. In this case, the ZP source size and brightness temperature
depend on the loop cross-section size. Furthermore, using a 3D
particle-in-cell model with the loss-cone type of distribution of hot electrons,
we compute the energy density of the upper-hybrid waves. Then this energy
density is compared with that of the electromagnetic waves and thus the
efficiency of the wave transformation (which is not well known) is estimated.

\begin{figure}
\centering
\includegraphics[width=6cm]{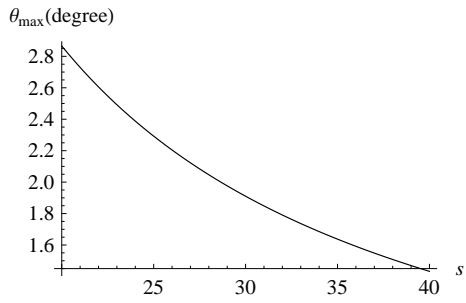}
\caption{Maximum escape angle of the plasma emission in conditions of the double-plasma resonance
depending on the gyro-harmonic number $s$.}
\label{figure1}
\end{figure}

\section{Sizes of the Zebra Source}

If in a plasma there is a source emitting in all directions then the emission
during its escape process obtains directional character. The range of angles
$\leq \Theta_\mathrm{max}$, in which the emission escapes, can be expressed as
(\opencite{1997riap.book.....Z})
\begin{eqnarray}
\Theta_\mathrm{max} = {\rm arcsec} \left(\frac{\omega}{\omega_{\rm L}}\right),
\label{eq1}
\end{eqnarray}
where $\omega_{\rm L}$ is the plasma frequency in the source and $\omega$ is
the emission frequency. In the conditions of a double-plasma resonance, the ratio
of these frequencies is (\opencite{2015A&A...581A.115K})
\begin{eqnarray}
\frac{\omega}{\omega_{\rm L}} = \frac{s}{\sqrt{s^2 - 1}},
\end{eqnarray}
where $s$ is the gyro-harmonic number.

In Figure~\ref{figure1} the maximum escape angle of the plasma emission for the
double-plasma resonance in the dependance on the gyro-harmonic number $s$ is shown.

Owing to the high directivity of the plasma emission (for emission frequency
close to the plasma frequency), the region from which the emission escapes can
be very small. This is connected with the fact that the emission region in the
flare loop at a fixed frequency is not flat, due to the density inhomogeneity
across the loop (the maximum density is expected at the loop axis). It has a
convex form and thus the emission with a high directivity (having the maximum
value in the direction perpendicular to the constant-density layer) can reach
an observer only from a limited region.

\begin{figure}
\centering
\includegraphics[width=6cm]{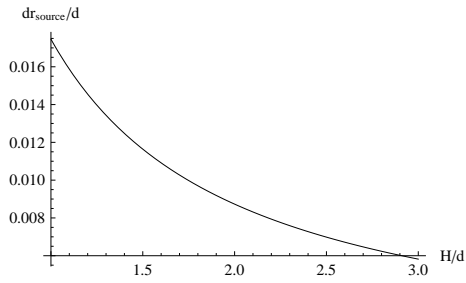}
\caption{The source extent [$\Delta r_\mathrm{source}/d$] as a function of $H/d$ for
the emission nearly perpendicular (for the maximum escape angle = 2$^\circ$)
to the constant-density profile.}
\label{figure2}
\end{figure}

\inlinecite{2000SoPh..193...77W} showed that the width of loops close to their
footpoints, where the decimetric bursts are generated, is about 0.5 arcsec
(0.36 Mm), and the typical width of higher loops is about 1 arcsec. Note that
the decimetric ZPs are generated in loops at low heights (about 3 Mm)
(\opencite{2015A&A...581A.115K};\opencite{2016SoPh..291.2037Y}). The width of
some loops can be even smaller.  For example,~\inlinecite{2013A&A...556A.104P}
found tiny 1.5 MK loop-like structures that they interpret as miniature coronal
loops. Their coronal segments above the chromosphere have a length of only
about 1 Mm and a thickness of less than 200 km.
Moreover,~\inlinecite{2012A&A...548A...1P} showed that in a 3D self-consistent
magnetohydrodynamic model of the solar corona, the loop width remains constant
with height, and profiles of intensities along the loop radius correspond to
gaussian ones. The gaussian profile along the loop radius was considered also
by~\inlinecite{1994SoPh..155..373C}. \inlinecite{2015SoPh..290...79K} assumed
the gaussian function ($\exp (-r^2/a^2)$, where $r$ is the loop radius and $a$
= 1 arcsec) describing the electron distribution in the flare loop.

Therefore, in agreement with the above-mentioned articles, we assume that the
density inside the magnetic loop at a specific height [$h_0$] has the
exponential form
\begin{eqnarray}
n_{\rm e}(r) = n_{\rm em} (h_0) \exp \left(-\frac{r^2}{d^2}\right),
\end{eqnarray}
where $n_{\rm em}$ is the density at the loop axis, $d$ is the loop width, and
$r$ is the radius across the loop. Moreover, the density in the loop decreases
with the height as $\approx \exp(- (h-h_0)/H$, where $h$ is the height in the
solar atmosphere and H is the scale height. Thus, the density inside the loop
can be expressed as
\begin{eqnarray}
n_{\rm e}(r, h) = n_{\rm em} (h_0) \exp \left(-\frac{r^2}{d^2}\right) \exp\left(- \frac{h-h_0}{H}\right).
\end{eqnarray}

Now, let us calculate the form of a layer with constant plasma density.
For the height where this layer is located, we can write
\begin{eqnarray}
C = n_{\rm em} (h_0) \exp \left(-\frac{r^2}{d^2}\right) \exp\left(- \frac{h-h_0}{H}\right),\\
h (r) = h_0 - H \frac{r^2}{d^2} + H \ln \left(\frac{n_{\rm em}(h_0)}{C} \right),
\end{eqnarray}
where $C$ is a constant.

In a loop with  constant magnetic field, from the pressure equilibrium and
density variations it follows that the temperature varies, and thus also the
scale-height. However, for simplification in further calculations, we assume
that the scale-height [$H$] inside the loop is constant.

Then the derivation of $\mathrm{d} h/ \mathrm{d} r$ is
\begin{equation}
\frac{\mathrm{d} h}{\mathrm{d} r} = -\frac{2 H r}{d^2}.
\label{der}
\end{equation}

Using this derivation, the extent of the emission region for which the emission
direction is nearly perpendicular to the constant-density profile can be
estimated as
\begin{equation}
\Delta r_{\rm source} = \frac{d^2 \tan(\Theta_\mathrm{max})}{2 H},
\label{ext}
\end{equation}
where $\Theta_\mathrm{max}$ is the maximum escape angle of the plasma emission
according to the Equation~\ref{eq1} (Figure~\ref{figure1}). Note that the
extent is independent of $r$. An example of the dependance of $\Delta r_{\rm
source}/d$ on $H/d$ for $\Theta_\mathrm{max}$ = 2$^\circ$ degrees is shown in
Figure~\ref{figure2}.

The emission source is also extended in height. This dimension can be estimated
as $h_s$ = $H(\mathrm{d}f/f)$, where d$f$ is the bandwidth of the zebra stripe and $f$ the
zebra-stripe frequency.

\begin{figure}
\centering
\includegraphics[width=10cm]{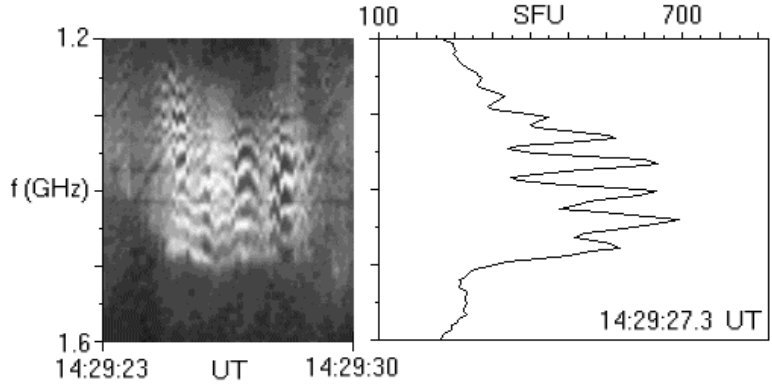}
\caption{\textit{Left panel}: An example of the zebra pattern observed by the Ond\v{r}ejov
radiospectrograph during the 2 May 1998 solar flare.
\textit{Right panel}: The radio-flux profile as a function of frequency at 14:29:27.3 UT.}
\label{figure3}
\end{figure}

\begin{figure}
\centering
\includegraphics[width=10cm]{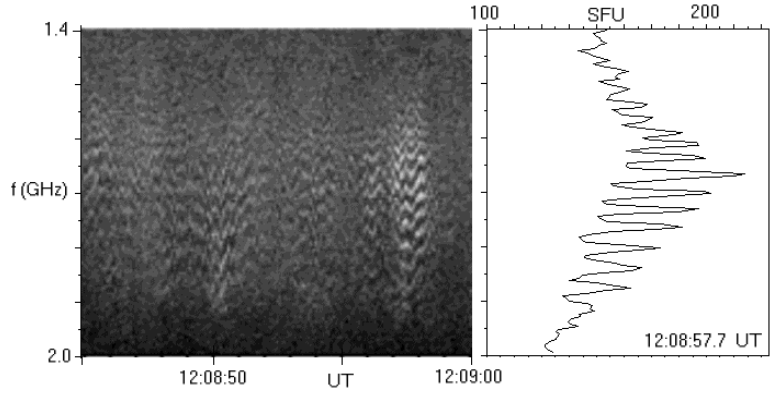}
\caption{\textit{Left panel}: An example of the zebra pattern observed by the Ond\v{r}ejov
radiospectrograph during the 14 February 1999 solar flare.
\textit{Right panel}: The radio-flux profile as a function of frequency at 12:08:57.7 UT.}
\label{figure4}
\end{figure}

\begin{figure}
\centering
\includegraphics[width=10cm]{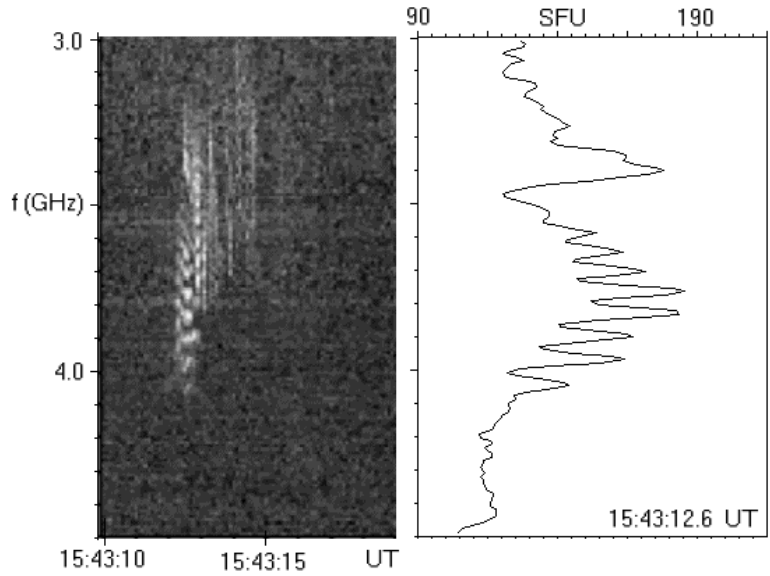}
\caption{\textit{Left panel}: An example of the zebra pattern observed by the Ond\v{r}ejov
radiospectrograph  during the 6 June 2000 solar flare.
\textit{Right panel}: The radio-flux profile as a function of frequency at 15:43:12.6 UT.}
\label{figure5}
\end{figure}

\section{Estimations of the Brightness Temperature of Zebra Structures}

Now, let us estimate the brightness temperature of some observed ZPs. For this
purpose we selected three ZPs observed by the Ond\v{r}ejov radiospectrograph
(\opencite{1993SoPh..147..203J}); see Figures~\ref{figure3},~\ref{figure4},
and~\ref{figure5}.

The brightness temperature can be expressed as (\opencite{1983SoPh...88..297Z})
\begin{equation}
T_b = \frac{S}{7 \times 10^{-11}} \frac{1}{f^2_{\rm GHz} L_8^2},
\label{bri}
\end{equation}
where $S$ is the radio flux in SFU, $f_{\rm GHz}$ is the frequency in GHz, and
$L_8$ (=  $ 2 \Delta r_{\rm source}$) is the dimension of the emission region
in units of 10$^8$~cm.

Thus to compute the brightness temperature, we need to determine the source
size [$\Delta r_{\rm source}$]. First, using the method presented by
\inlinecite{2015A&A...581A.115K}, we determined the gyro-harmonic numbers $s_1$
for the observed zebras. Knowing $s_1$ (see Table~\ref{tab1}) and considering
the scale height as $H = 1$~Mm (according to the relation $H$~[m]$ = 50\,T$ [K]
(\opencite{2014masu.book.....P}) for the temperature $T = 2 \times 10^{4}$
K), we calculated the source size of the observed zebras for two values of $2
d$ (1 and 2 arcsec). All of the computed parameters of the zebra sources together with
the brightness temperatures are shown in Table~\ref{tab1}.

\begin{table}
\caption{ZP source parameters. $S$ is the radio flux in SFU units and $s_1$ is
the gyro-number of the stripe with the lowest frequency.} \centering
\begin{tabular}{cccc}
\hline
 & ZP 2 May 1998  & ZP 14 February 1999 & ZP 6 June 2000  \\
\hline
Event location & S15W15 & N16E09 & N23E15 \\
$S\,[ SFU ] $    & 650    & 170  & 210 \\
$f_{\rm GHz}$    & 1.45    & 1.67  & 3.78    \\
$s_1$    & 21    & 32  & 34    \\
$\Theta_\mathrm{max}$    & 2.70    & 1.79  & 1.68 \\
$L_8$ (2 $d=1$ arcsec) & 0.0059 & 0.0038 & 0.0035 \\
$L_8$ (2 $d=2$ arcsec) &0.023   &0.015   &0.014   \\
$T_b$ (2 $d =1$ arcsec) &13 $\times$ 10$^{16}$ K&6 $\times$ 10$^{16}$ K&1.7 $\times$ 10$^{16}$ K      \\
$T_b$ (2 $d =2$ arcsec) &0.83 $\times$ 10$^{16}$ K& 0.39 $\times$ 10$^{16}$ K&0.11 $\times$ 10$^{16}$ K      \\
Source height [km]    &  28   &  14.7    &  12.5  \\
\hline \label{tab1}
\end{tabular}
\end{table}

However, in a region with a rapid change of the plasma temperature, the scale
height can be shorter. Therefore, let us estimate the brightness temperature
using the model by \inlinecite{2008A&A...488.1079S}. For typical densities in
ZP sources 5.0$\times$ 10 $^ 9$ cm $^{-3}$ -- 3.6$\times$ 10$ ^ {10}$ cm$^{-3}$
the model gives heights in the solar atmosphere between 2.84 Mm and 3.27 Mm.
Thus, the scale height  is $H$ = 0.21~Mm, which is almost five times shorter
than that according to the formula of Priest used above. For such a scale
height the ZP source parameters are given in Table~\ref{tab2}.

\begin{table}
\caption{ZP source parameters for $H$=0.21Mm.} \centering
\begin{tabular}{cccc}
\hline
 & ZP 2 May 1998  & ZP 14 February 1999 & ZP 6 June 2000  \\
\hline
$L_8$ (2 $d=1$ arcsec) & 0.028 & 0.018 & 0.017 \\
$L_8$ (2 $d=2$ arcsec) &0.11   &0.070   &0.067   \\
$T_b$ (2 $d=1$ arcsec) &5.6 $\times$ 10$^{15}$ K&2.7 $\times$ 10$^{15}$ K&0.73 $\times$ 10$^{15}$ K      \\
$T_b$ (2 $d=2$ arcsec) &0.36 $\times$ 10$^{15}$ K& 0.18 $\times$ 10$^{15}$ K&0.047 $\times$ 10$^{15}$ K      \\
\hline \label{tab2}
\end{tabular}
\end{table}

\section{Energy Densities of Electromagnetic and Upper-Hybrid Waves in the ZP Source}

Let us consider the zebra observed during the 2 May 1998 event. Knowing the ZP
radio flux (650 SFU) and the zebra line frequency width (40 MHz), and computing
the ratio of the emission area at 1 AU (corresponding to the emission
directivity angle (2.7$^\circ$ for $s=21$, see Figure~\ref{figure1})), and ZP
source area ($L_8 \times L_8$ for four cases according to Tables~\ref{tab1}
and~\ref{tab2}), the energy density of electromagnetic waves in the ZP source
is calculated; see the second column in Table~\ref{tab3}.

In the double plasma resonance (DPR) model of ZPs, it is assumed that in the ZP
source there are hot electrons with a loss-cone type distribution together with
much denser background plasma. The distribution is unstable and generates the
upper-hybrid waves that after their transformation produce the observed ZPs.

Therefore, besides the estimation of the energy density of electromagnetic
waves, it is highly desirable to estimate the energy density of the
upper-hybrid waves in the ZP source. For this purpose we use a 3D
particle-in-cell (PIC) relativistic code
(\opencite{1985stan.reptR....B},~\opencite{matsumoto.1993},~\opencite{2008SoPh..247..335K}).
Although this code is very useful for such computations, it has its own
limitations. Therefore some parameters of the ZP of 2 May 1998 cannot be
reproduced in the following PIC computations. For example, there is a problem
in making computations with high gyro-harmonic numbers ($s>20$ in our case)
because it is very difficult to select PIC plasma parameters reproducing
resonances with high-harmonic numbers, especially due to the discrete space
(grids) in PIC models.

The size of the model is $L_\mathrm{x} \times L_\mathrm{y} \times L_\mathrm{z}$
= $\lambda \Delta \times \lambda \Delta \times 32 \Delta$, where $\Delta$ is
the grid size and $\lambda$ is the wavelength of the upper-hybrid wave in
resonance in normalized units. We chose a model containing only one wavelength
of the upper-hybrid wave to simplify the processing and decrease computing
time. The model time step is $\mathrm{d}t = 1$, plasma electron frequency
$\omega_\mathrm{pe} \mathrm{d}t = 0.05$, initial magnetic field is in
$z$-direction, electron-cyclotron frequency is, e.g., $\omega_\mathrm{ce} =
0.176~\omega_\mathrm{pe}$ for the the harmonic number $s=7$. The harmonic
number is considered in the interval $s$ = 4 -- 18. We use two groups of
electrons: a) cold background electrons with the thermal velocity
$v_{\mathrm{tb}} = 0.03$~c (c is the light speed), corresponding to temperature
5.35~MK, and b) hot electrons with the DGH (\opencite{1965PhRvL..14..131D})
distribution function for $j=1$ in the form
\begin{equation}
f = \frac{u_\perp^2}{2 (2\pi)^{3/2} v_\mathrm{t}^5} \exp \left(-\frac{u_\perp^2 + u_\parallel^2}{2 v_\mathrm{t}^2}\right),
\label{edory}
\end{equation}
where $u_\perp = p_\perp/m_\mathrm{e}$ and $u_\parallel =
p_\parallel/m_\mathrm{e}$ are electron velocities and $p_\perp$ and
$p_\parallel$ are components of the electron momentum perpendicular and
parallel to the magnetic field, $m_e$ is the electron mass, and $v_\mathrm{t} =
0.2$~c is the thermal spread in the velocities of hot electrons.

The electron density of cold electrons per cell was taken as $n_\mathrm{e} =
1920$ and the ratio of hot and background plasma electrons as
$n_\mathrm{h}/n_\mathrm{e} = 1/8$. We also made computations with
$n_\mathrm{h}/n_\mathrm{e} =$ 1/16 and 1/32 to know the dependence of the
saturation energy of the upper-hybrid waves on the density ratio
$n_\mathrm{h}/n_\mathrm{e}$. The number of protons is the same as electrons and
their temperature is the same as that of cold electrons.

\begin{figure}
\centering
\includegraphics[width=6cm]{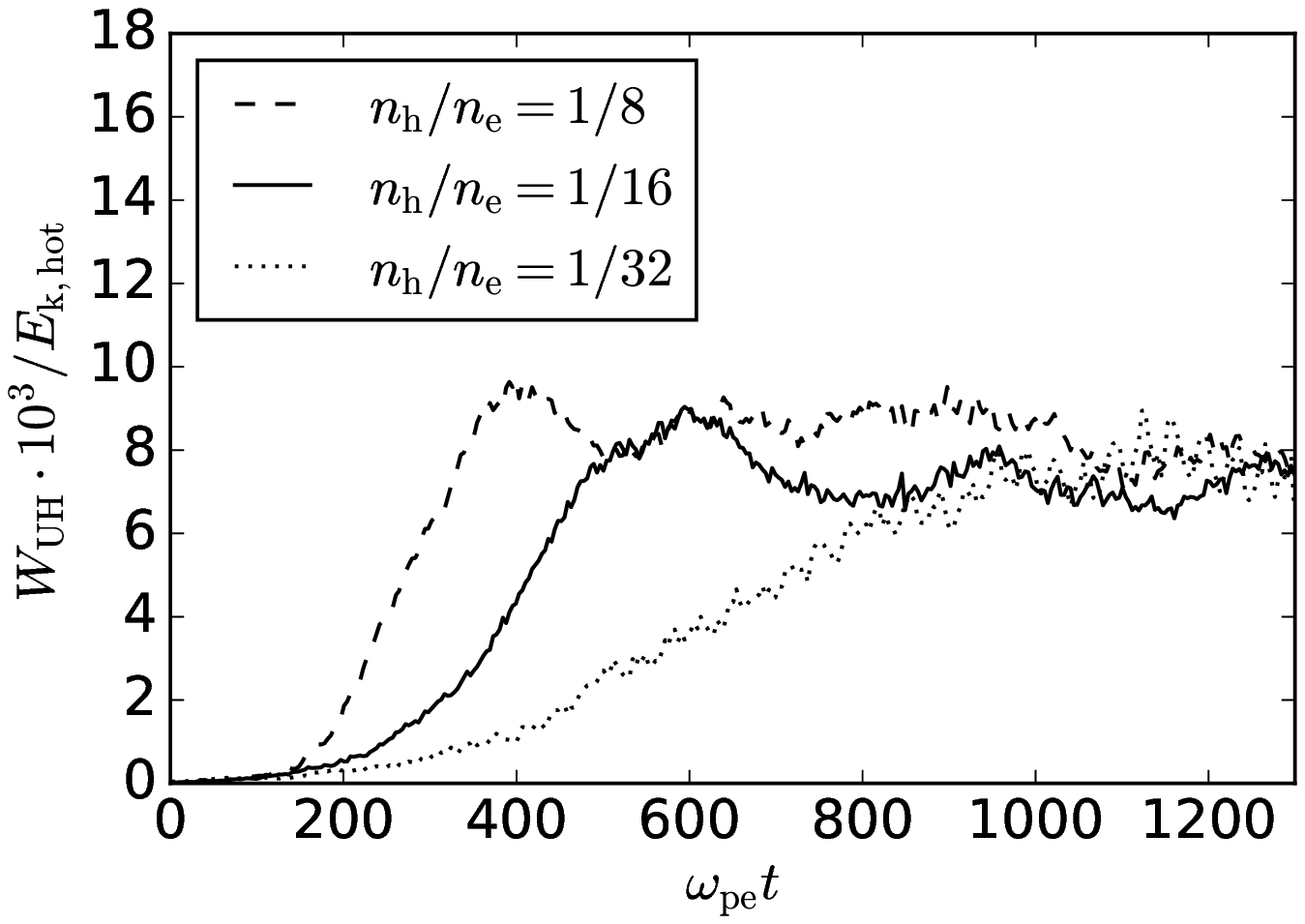}
\includegraphics[width=6cm]{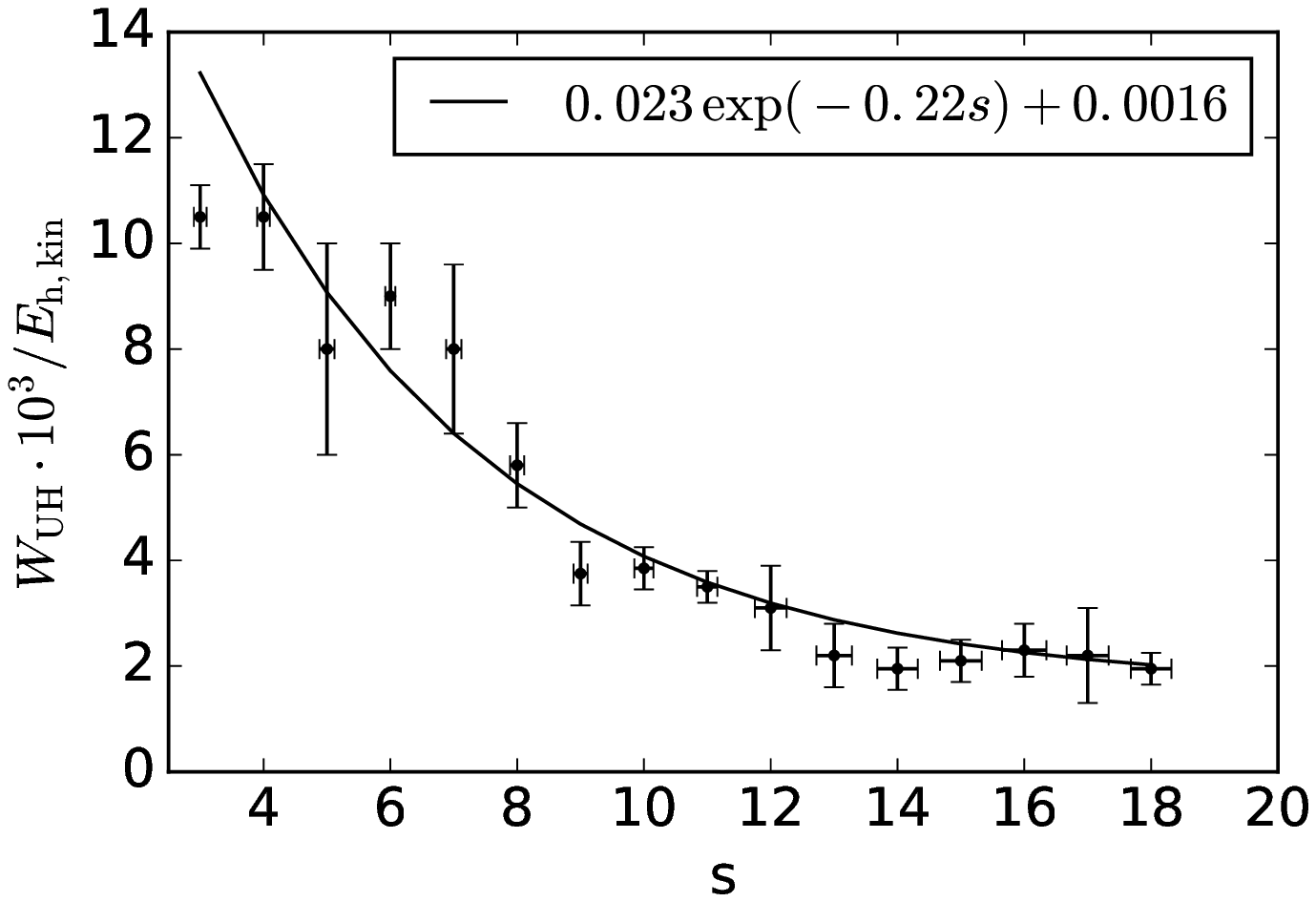}
\caption{\textit{Left: }Temporal evolution of the ratio of the
energy of the upper-hybrid waves $W_\mathrm{UH}$ to the kinetic energy of hot electrons $E_\mathrm{h,kin}$
for s = 7 and three values of $n_\mathrm{h}/n_\mathrm{e}$. \textit{Right: }The ratio of the saturated energy
of the upper-hybrid waves to the kinetic energy of hot electrons for $n_\mathrm{h}/n_\mathrm{e}$ = 1/8
as a function of $s$. The full line shows the exponential fit.}
\label{figure6}
\end{figure}

\begin{table}
\caption{Energy densities of electromagnetic and upper-hybrid waves in the ZP
source for the 2 May 1998 event. $\epsilon = W_\mathrm{elm}/W_\mathrm{UH,min}$ is
the parameter expressing the efficiency of transformation of the upper-hybrid
waves into electromagnetic ones.} \centering
\begin{tabular}{cccc}
\hline
$L_8$ & $W_\mathrm{elm}$  & $W_\mathrm{UH,min}$ & $\epsilon$ \\
   $\left[\mathrm{Mm}\right]$ &   $\left[\mathrm{J~m}^{-3}\right]$  &  $\left[\mathrm{J~m}^{-3}\right]$  &          \\
\hline
0.0059 & 3.90$\times$ 10$^{-8}$ & 4.40$\times$ 10$^{-5}$ & 8.86$\times$ 10$^{-4}$\\
0.023  & 2.57$\times$ 10$^{-9}$  & 4.40$\times$ 10$^{-5}$ &   5.84$\times$ 10$^{-5}$\\
0.028 & 1.73$\times$ 10$^{-9}$ & 4.40$\times$ 10$^{-5}$ & 3.93$\times$ 10$^{-5}$\\
0.11  &  1.12$\times$ 10$^{-10}$ & 4.40$\times$ 10$^{-5}$ &  2.54$\times$ 10$^{-6}$\\
\hline \label{tab3}
\end{tabular}
\end{table}

Using this PIC code, the temporal evolution of the energy of the upper-hybrid
waves [$W_\mathrm{UH}$], generated by hot electrons, for $s=7$ and three values
of $n_\mathrm{h}/n_\mathrm{e}$ are shown in Figure~\ref{figure6} left. As can
be seen here, the ratio of the saturated energy of the upper-hybrid waves to
the kinetic energy of hot electrons is in all three cases the same
($W_\mathrm{UH}$ = 8 $\times$ 10$^{-3}$ $E_\mathrm{h, kin}$).  This means that
the saturated energy of the upper-hybrid waves is proportional to the
$n_\mathrm{h}/n_\mathrm{e}$ ratio because $E_\mathrm{h, kin}$ depends
on~$n_\mathrm{h}$. The saturated energy of the upper-hybrid waves also depends
on $s$ as shown in Figure~\ref{figure6} right. The computed values can be well
fitted by the exponential fit. Therefore, for the 2 May 1998 zebra that was
analyzed, where $s=21$, we use this exponential fit, which gives the value of
the saturated energy of the upper-hybrid waves as $W_\mathrm{UH}$ =
1.6$\times$ 10$^{-3}$ $E_\mathrm{h, kin}$, where $E_\mathrm{h, kin}$
is the kinetic energy of hot electrons.

Now, for the following estimations, let us derive the minimum value of the
parameter $n_\mathrm{h}/n_\mathrm{e}$. For this purpose we used the analytical
expression for the growth rate of the upper-hybrid waves as derived by
\inlinecite{1991SoPh..132..173T}
\begin{equation}
 - \gamma_\mathrm{UH} \approx 4.4 \times 10^{-2} \omega_\mathrm{pe} \frac{n_\mathrm{h}}{n_\mathrm{e}}.
\end{equation}
This growth rate agrees with that in our PIC simulations. To generate the
upper-hybrid waves this growth rate needs to be greater than the damping of
these waves by collisions,
\begin{equation}
\gamma_\mathrm{c} = 2.75 \frac{n_\mathrm{e}}{T_\mathrm{e}^{3/2}} \ln \left(10^4 T_\mathrm{e}^{3/2}/n_\mathrm{e}^{1/3}\right),
\end{equation}
where $T_\mathrm{e}$ is the background plasma temperature. Thus, when we put
$\gamma_\mathrm{UH}$ equal to $\gamma_\mathrm{c}$, then for the mean ZP
frequency (1.45 GHz and corresponding plasma density $n_\mathrm{e}$ = 2.6
$\times$ 10$^{16}$ m$^{-3}$) and for $T_\mathrm{e}$ in the ZP source as 2 $\times$
10$^{4}$ K (at bottom of the transition region), the minimal ratio of
$n_\mathrm{h}/n_\mathrm{e}$ is 4.93 $\times$ 10$^{-4}$.

Now, if we take the density of the hot electrons in the ZP source as equal to
the minimum density $n_\mathrm{h}$ =  2.6 $\times$ 10$^{16}$ $\times$ 4.93
$\times$ 10$^{-4}$ = 1.28 $\times$ 10$^{13}$~m$^{-3}$ and utilizing the
extrapolated value of the saturated energy of the upper-hybrid waves for $s = 21$
and the linear dependance of the saturated energy of the upper-hybrid waves on
$n_\mathrm{h}$, found in the PIC simulations, then the minimum energy density
of the upper-hybrid waves $W_\mathrm{UH,min}$ is calculated; see the third
column in Table~\ref{tab3}.

Then, if we assume that the energy density of the upper-hybrid waves in the ZP
source from 2 May 1998 is the same as $W_\mathrm{UH,min}$, then we can
calculate the parameter expressing the efficiency of transformation of the
upper-hybrid waves into electromagnetic ones; see the last column in
Table~\ref{tab3}. However, the ratio $n_\mathrm{h}/n_\mathrm{e}$ in real
conditions needs to be greater than its minimum value, and therefore $\epsilon$
in real conditions should be lower. On the other hand, values of the parameter
$\epsilon$ would be greater if we considered the absorption of the
electromagnetic waves near the ZP source.

\section{Discussion and Conclusions}

The closeness of the emission frequency of decimetric zebras to the plasma
frequency determines a narrow directivity of the ZP emission. For the
exponential density profile across the flaring loop, it gives a small area of
such emission and thus high brightness temperatures. We considered two variants
of the density dependance on height, {\it i.e.}, two values of the scale
height: 1 Mm according to the formula of \inlinecite{2014masu.book.....P} and
0.21 Mm for the transition region (\opencite{2008A&A...488.1079S}), and two
values for the loop width (1 and 2 arcseconds). In all cases high brightness
temperatures were obtained. For the higher value of the density scale height
the brightness temperature was estimated as 1.1 $\times$ 10$^{15}$~--~1.3
$\times$ 10$^{17}$ K, and for the lower value it was estimated as 4.7 $\times$
10$^{13}$ -- 5.6 $\times$ 10$^{15}$.

As shown in the Introduction, previous estimations of the ZP brightness
temperature were noticeably lower (from 2 $\times$ 10$^7$ to 10$^{13}$ K). The
high brightness temperature found here together with short duration of zebras
and their frequent strong polarization can only be explained as generated by
the coherent emission mechanism. Namely, in the non-coherent emission
mechanism, the brightness temperature cannot be higher than 10$^{12}$ K, which
is given by the Compton limit. The mechanism of the coherent emission of the
plasma waves (including the upper-hybrid waves considered here) is described in
detail, e.g. by \inlinecite{1998PhyU...41.1157F}. Here, we only mention that
for the processes described in the present article an anisotropic distribution
of superthermal electrons is necessary. As shown by
\inlinecite{2015SoPh..290.2001Y} the most probable location of the zebra
generation is the transition region in the solar atmosphere of active regions.
In the transition region the temperature as well as the pressure changes
rapidly, and thus the magnetic fields fan out to form funnels; see, e.g.
\inlinecite{2014A&ARv..22...78W}. In such magnetic field funnels superthermal
electrons with momentum perpendicular to the magnetic field are more numerous
than those with parallel momentum. The consequence is that in this spatially
small region the high level of anisotropy of the superthermal electrons is
easily reached, and thus the upper-hybrid waves are generated there
(\opencite{2017A&A...555A...1B}).

Note that these brightness temperatures are close to the brightness
temperatures of decimetric spikes (\opencite{1986SoPh..104...99B}), which
indicates that energies of electrons in the two types of bursts are
similar. Because the emission frequency is close to the plasma frequency, the
emission absorption can be high and thus the brightness temperature can be even
higher.

Observed sizes of ZP sources can be larger than those assumed here because they
are enlarged by the scattering of the emission on density fluctuations in the
corona (\opencite{1994ApJ...426..774B}).

We found a lower brightness temperature for a shorter scale height. It is
indicated independently in findings presented by
\inlinecite{2015SoPh..290.2001Y}, \inlinecite{2015A&A...581A.115K} and
\inlinecite{2016SoPh..291.2037Y} that the observational probability of a burst
with zebras, which is generated in the transition region with a steep density
gradient, is generally greater than the burst generated in the region with
smoother changes of the plasma density. This is caused by an enlargement of the
visible emission area in the atmosphere with the high density gradient.

Note that sometimes ZPs appear on the radio spectrum of a Type IV (continuum)
burst irregularly or quasi-periodically (on timescales of seconds). It can be
explained by small irregular or quasi-periodic motions of the flare loop in the
case when the ZP source area is sufficiently small. Then the narrow cone of the
emission directivity is oriented toward an observer and the zebra is observed,
or vice versa.

Numerical simulations with the 3D particle-in-cell model having hot electrons
described by the DGH distribution function show that firstly the energy density
of the upper-hybrid waves exponentially grows with the linear growth rate and
then it is saturated. We found that the saturation energy of the upper-hybrid
waves is proportional to the ratio $n_\mathrm{h}/n_\mathrm{e}$. This dependance
enabled us to calculate the saturation energy of the upper-hybrid waves for
much smaller ratios of $n_\mathrm{h}/n_\mathrm{e}$. The saturation energy of
the upper-hybrid waves also depends on the gyro-harmonic number $s$. For $s =
7-18$ we found that the computed saturated energies can be well fitted by a
exponential function. This fit enables us to find the saturated energy of the
upper-hybrid waves for the analyzed 2 May 1998 zebra with $s = 21$ as about
$W_\mathrm{UH}$= 1.6 $\times$ 10$^{-3}$ $E_\mathrm{h,kin}$.

The upper-hybrid waves are generated when the growth rate exceeds the
damping of these waves by collisions. For the zebra observed during the 2 May
1998 event this condition is fulfilled if the ratio of hot and cold electrons
$n_\mathrm{h}/n_\mathrm{e}$ is greater than 4.93 $\times$ 10$^{-4}$. Using this
value we computed the minimum energy density of the upper-hybrid waves in ZP
source ($W_\mathrm{UH,min}$ = 4.40 $\times$ 10$^{-5}$ J m$^{-3}$) and the
transformation efficiency of the upper-hybrid waves into electromagnetic ones
($\epsilon$ = 2.54 $\times$ 10$^{-6}$ -- 8.86 $\times$ 10$^{-4}$).

The transformation efficiency strongly depends on plasma parameters in
the radio source such as plasma turbulence, levels of low-frequency plasma waves
(\textit{e.g.} the ion-sound waves), and density gradients \cite{1985srph.book..177M}.
Unfortunately, most of these parameters in ZP sources are not known, and moreover
the theory of wave conversions is not fully established, especially in
the non-linear regime. For example, \inlinecite{1985srph.book..177M} presents this
efficiency in the very broad range from 10$^{-10}$ for the scattering on the
thermal ions to 10$^{-4}$ for the small-scale density inhomogeneities (10 --
10$^{2}$ km). Comparing now the transformation efficiency found in the present
article with those shown by \inlinecite{1985srph.book..177M}, we think that in the
zebra source there are small-scale density inhomogeneities.

%--------------------------------------------------------------------------------------------------------------------------------------------------------
%acknowledgements

\begin{acks}
The authors thank the referee for constructive comments that improved the
article. M. Karlick\'y acknowledges support from Grants 16-13277S and 17-16447S
of the Grant Agency of the Czech Republic. L.V. Yasnov acknowledge support from
Grant 16-02-00254  of the Russian Foundation for Basic Research. Computational
resources were provided by the CESNET LM2015042 and the CERIT Scientific Cloud
LM2015085, provided under the programme ``Projects of Large Research,
Development, and Innovations Infrastructures``
\end{acks}

\section*{Disclosure of Potential Conflicts of Interest}
Authors have no potential conflict of interest.

%--------------------------------------------------------------------------------------------------------------------------------------------------------
% Bibliography
% Using BibTeX
%
\bibliographystyle{spr-mp-sola}
% %\bibliographystyle{spr-mp-sola-cnd} %% Alternative style: no title, no concluding page
\bibliography{L2017a}

\begin{thebibliography}{36}
% BibTex style file: spr-mp-sola.bst (nameyear), 2015-03-09
\ifx\bisbn     \undefined \def\bisbn  #1{ISBN #1}\fi
\ifx\binits    \undefined \def\binits#1{#1}\fi
\ifx\bauthor   \undefined \def\bauthor#1{#1}\fi
\ifx\batitle   \undefined \def\batitle#1{#1}\fi
\ifx\bjtitle   \undefined \def\bjtitle#1{\textit{#1}}\fi
\ifx\bvolume   \undefined \def\bvolume#1{\textbf{#1}}\fi
\ifx\byear     \undefined \def\byear#1{#1}\fi
\ifx\bissue    \undefined \def\bissue#1{#1}\fi
\ifx\bfpage    \undefined \def\bfpage#1{#1}\fi
\ifx\blpage    \undefined \def\blpage #1{#1}\fi
\ifx\burl      \undefined \def\burl#1{\textsf{#1}}\fi
\ifx\href      \undefined \def\href#1#2{\textsf{#2}}\fi
\ifx\betal     \undefined \def\betal{\textit{et al.}}\fi
\ifx\bctitle   \undefined \def\bctitle#1{#1}\fi
\ifx\beditor   \undefined \def\beditor#1{#1}\fi
\ifx\bbtitle   \undefined \def\bbtitle#1{\textit{#1}}\fi
\ifx\bedition  \undefined \def\bedition#1{#1}\fi
\ifx\bseriesno \undefined \def\bseriesno#1{\textbf{#1}}\fi
\ifx\blocation \undefined \def\blocation#1{#1}\fi
\ifx\bsertitle \undefined \def\bsertitle#1{\textit{#1}}\fi
\ifx\bsnm      \undefined \def\bsnm#1{#1}\fi
\ifx\bsuffix   \undefined \def\bsuffix#1{#1}\fi
\ifx\bparticle \undefined \def\bparticle#1{#1}\fi
\ifx\barticle  \undefined \def\barticle#1{}\fi
\ifx\binstitute  \undefined \def\binstitute#1{#1}\fi
\ifx\bpublisher  \undefined \def\bpublisher#1{#1}\fi
\ifx\doiurl    \undefined
  \def\doiurl#1{\href{http://dx.doi.org/#1}{\textsf{DOI}}}\fi
\ifx\arxivurl  \undefined
  \def\arxivurl#1{\href{http://arxiv.org/abs/#1}{\textsf{arXiv}}}\fi
\ifx\adsurl    \undefined
  \def\adsurl#1{\href{http://adsabs.harvard.edu/abs/#1}{\textsf{ADS}}}\fi
\ifx\botherref \undefined \def\botherref#1{}\fi
\ifx\url       \undefined \def\url#1{\textsf{#1}}\fi
\ifx\bchapter  \undefined \def\bchapter#1{}\fi
\ifx\bbook     \undefined \def\bbook#1{}\fi
\ifx\bcomment  \undefined \def\bcomment#1{#1}\fi
\ifx\oauthor   \undefined \def\oauthor#1{#1}\fi
\ifx\citeauthoryear \undefined\def \citeauthoryear#1{#1}\fi
\ifx\endbibitem\undefined \def\endbibitem{}\fi
\ifx\bconflocation  \undefined \def\bconflocation#1{#1} \fi

\bibitem[\protect\citeauthoryear{{Altyntsev}
  \textit{et~al.}}{2005}]{2005A&A...431.1037A}
\begin{barticle}
\bauthor{\bsnm{{Altyntsev}}, \binits{A.T.}},
\bauthor{\bsnm{{Kuznetsov}}, \binits{A.A.}},
\bauthor{\bsnm{{Meshalkina}}, \binits{N.S.}},
\bauthor{\bsnm{{Rudenko}}, \binits{G.V.}},
\bauthor{\bsnm{{Yan}}, \binits{Y.}}:
\byear{2005},
\batitle{{On the origin of microwave zebra pattern}}.
\bjtitle{\aap}
\bvolume{431},
\bfpage{1037}.
\doiurl{10.1051/0004-6361:20048337}.
\adsurl{http://cdsads.u-strasbg.fr/abs/2005A\%26A...431.1037A}.
\end{barticle}
\endbibitem

\bibitem[\protect\citeauthoryear{{B{\'a}rta} and
  {Karlick{\'y}}}{2006}]{2006A&A...450..359B}
\begin{barticle}
\bauthor{\bsnm{{B{\'a}rta}}, \binits{M.}},
\bauthor{\bsnm{{Karlick{\'y}}}, \binits{M.}}:
\byear{2006},
\batitle{{Interference patterns in solar radio spectra: High-resolution
  structural analysis of the corona}}.
\bjtitle{\aap}
\bvolume{450},
\bfpage{359}.
\doiurl{10.1051/0004-6361:20054386}.
\adsurl{http://cdsads.u-strasbg.fr/abs/2006A\%26A...450..359B}.
\end{barticle}
\endbibitem

\bibitem[\protect\citeauthoryear{{Bastian}}{1994}]{1994ApJ...426..774B}
\begin{barticle}
\bauthor{\bsnm{{Bastian}}, \binits{T.S.}}:
\byear{1994},
\batitle{{Angular scattering of solar radio emission by coronal turbulence}}.
\bjtitle{\apj}
\bvolume{426},
\bfpage{774}.
\doiurl{10.1086/174114}.
\adsurl{http://cdsads.u-strasbg.fr/abs/1994ApJ...426..774B}.
\end{barticle}
\endbibitem

\bibitem[\protect\citeauthoryear{{Ben\'a\v{c}ek}, {Karlick\'y}, and
  {Yasnov}}{2017}]{2017A&A...555A...1B}
\begin{barticle}
\bauthor{\bsnm{{Ben\'a\v{c}ek}}, \binits{J.}},
\bauthor{\bsnm{{Karlick\'y}}, \binits{M.}},
\bauthor{\bsnm{{Yasnov}}, \binits{L.}}:
\byear{2017},
\batitle{{Temperature dependent growth rates of the upper-hybrid waves and
  solar radio zebra patternss}}.
\bjtitle{\aap}
\bvolume{555},
\bfpage{A1}.
\doiurl{10.1051/0004-6361/201219473}.
\adsurl{http://cdsads.u-strasbg.fr/abs/2012A\%26A...548A...1P}.
\end{barticle}
\endbibitem

\bibitem[\protect\citeauthoryear{{Benz}}{1986}]{1986SoPh..104...99B}
\begin{barticle}
\bauthor{\bsnm{{Benz}}, \binits{A.O.}}:
\byear{1986},
\batitle{{Millisecond radio spikes}}.
\bjtitle{\solphys}
\bvolume{104},
\bfpage{99}.
\doiurl{10.1007/BF00159950}.
\adsurl{http://cdsads.u-strasbg.fr/abs/1986SoPh..104...99B}.
\end{barticle}
\endbibitem

\bibitem[\protect\citeauthoryear{{Buneman} and
  {Storey}}{1985}]{1985stan.reptR....B}
\begin{botherref}
\oauthor{\bsnm{{Buneman}}, \binits{O.}},
\oauthor{\bsnm{{Storey}}, \binits{L.R.O.}}:
1985,
{Simulations of fusion plasmas by A 3-D, E-M particle code}.
Technical report,
Stanford Univ. Report,
Stanford.
\adsurl{http://cdsads.u-strasbg.fr/abs/1985stan.reptR....B}.
\end{botherref}
\endbibitem

\bibitem[\protect\citeauthoryear{{Chen}
  \textit{et~al.}}{2011}]{2011ApJ...736...64C}
\begin{barticle}
\bauthor{\bsnm{{Chen}}, \binits{B.}},
\bauthor{\bsnm{{Bastian}}, \binits{T.S.}},
\bauthor{\bsnm{{Gary}}, \binits{D.E.}},
\bauthor{\bsnm{{Jing}}, \binits{J.}}:
\byear{2011},
\batitle{{Spatially and spectrally resolved observations of a zebra Pattern in
  a solar decimetric radio burst}}.
\bjtitle{\apj}
\bvolume{736},
\bfpage{64}.
\doiurl{10.1088/0004-637X/736/1/64}.
\adsurl{http://cdsads.u-strasbg.fr/abs/2011ApJ...736...64C}.
\end{barticle}
\endbibitem

\bibitem[\protect\citeauthoryear{{Chernov}}{2006}]{2006SSRv..127..195C}
\begin{barticle}
\bauthor{\bsnm{{Chernov}}, \binits{G.P.}}:
\byear{2006},
\batitle{{Solar Radio Bursts with Drifting Stripes in Emission and
  Absorption}}.
\bjtitle{\ssr}
\bvolume{127},
\bfpage{195}.
\doiurl{10.1007/s11214-006-9141-7}.
\adsurl{http://cdsads.u-strasbg.fr/abs/2006SSRv..127..195C}.
\end{barticle}
\endbibitem

\bibitem[\protect\citeauthoryear{{Chernov}, {Yan}, and
  {Fu}}{2003}]{2003A&A...406.1071C}
\begin{barticle}
\bauthor{\bsnm{{Chernov}}, \binits{G.P.}},
\bauthor{\bsnm{{Yan}}, \binits{Y.H.}},
\bauthor{\bsnm{{Fu}}, \binits{Q.J.}}:
\byear{2003},
\batitle{{A superfine structure in solar microwave bursts}}.
\bjtitle{\aap}
\bvolume{406},
\bfpage{1071}.
\doiurl{10.1051/0004-6361:20030779}.
\adsurl{http://cdsads.u-strasbg.fr/abs/2003A\%26A...406.1071C}.
\end{barticle}
\endbibitem

\bibitem[\protect\citeauthoryear{{Chernov}
  \textit{et~al.}}{1994}]{1994SoPh..155..373C}
\begin{barticle}
\bauthor{\bsnm{{Chernov}}, \binits{G.P.}},
\bauthor{\bsnm{{Klein}}, \binits{K.-L.}},
\bauthor{\bsnm{{Zlobec}}, \binits{P.}},
\bauthor{\bsnm{{Aurass}}, \binits{H.}}:
\byear{1994},
\batitle{{Fine structure in a metric type 4 burst: Multi-site spectrographic,
  polarimetric, and heliographic observations}}.
\bjtitle{\solphys}
\bvolume{155},
\bfpage{373}.
\doiurl{10.1007/BF00680601}.
\adsurl{http://cdsads.u-strasbg.fr/abs/1994SoPh..155..373C}.
\end{barticle}
\endbibitem

\bibitem[\protect\citeauthoryear{{Dory}, {Guest}, and
  {Harris}}{1965}]{1965PhRvL..14..131D}
\begin{barticle}
\bauthor{\bsnm{{Dory}}, \binits{R.A.}},
\bauthor{\bsnm{{Guest}}, \binits{G.E.}},
\bauthor{\bsnm{{Harris}}, \binits{E.G.}}:
\byear{1965},
\batitle{{Unstable Electrostatic Plasma Waves Propagating Perpendicular to a
  Magnetic Field}}.
\bjtitle{Physical Review Letters}
\bvolume{14},
\bfpage{131}.
\doiurl{10.1103/PhysRevLett.14.131}.
\adsurl{1965PhRvL..14..131D}.
\end{barticle}
\endbibitem

\bibitem[\protect\citeauthoryear{{Fleishman} and
  {Mel'nikov}}{1998}]{1998PhyU...41.1157F}
\begin{barticle}
\bauthor{\bsnm{{Fleishman}}, \binits{G.D.}},
\bauthor{\bsnm{{Mel'nikov}}, \binits{V.F.}}:
\byear{1998},
\batitle{{Reviews of Topical Problems: Millisecond solar radio spikes}}.
\bjtitle{PHYSICS USPEKHI}
\bvolume{41},
\bfpage{1157}.
\doiurl{10.1070/PU1998v041n12ABEH000510}.
\adsurl{http://cdsads.u-strasbg.fr/abs/1998PhyU...41.1157F}.
\end{barticle}
\endbibitem

\bibitem[\protect\citeauthoryear{{Jiricka}
  \textit{et~al.}}{1993}]{1993SoPh..147..203J}
\begin{barticle}
\bauthor{\bsnm{{Jiricka}}, \binits{K.}},
\bauthor{\bsnm{{Karlicky}}, \binits{M.}},
\bauthor{\bsnm{{Kepka}}, \binits{O.}},
\bauthor{\bsnm{{Tlamicha}}, \binits{A.}}:
\byear{1993},
\batitle{{Fast drift burst observations with the new Ondrejov
  radiospectrograph}}.
\bjtitle{\solphys}
\bvolume{147},
\bfpage{203}.
\doiurl{10.1007/BF00675495}.
\adsurl{http://cdsads.u-strasbg.fr/abs/1993SoPh..147..203J}.
\end{barticle}
\endbibitem

\bibitem[\protect\citeauthoryear{{Karlick{\'y}}}{2013}]{2013A&A...552A..90K}
\begin{barticle}
\bauthor{\bsnm{{Karlick{\'y}}}, \binits{M.}}:
\byear{2013},
\batitle{{Radio continua modulated by waves: Zebra patterns in solar and pulsar
  radio spectra?}}
\bjtitle{\aap}
\bvolume{552},
\bfpage{A90}.
\doiurl{10.1051/0004-6361/201321356}.
\adsurl{http://cdsads.u-strasbg.fr/abs/2013A\%26A...552A..90K}.
\end{barticle}
\endbibitem

\bibitem[\protect\citeauthoryear{{Karlick{\'y}} and
  {B{\'a}rta}}{2008}]{2008SoPh..247..335K}
\begin{barticle}
\bauthor{\bsnm{{Karlick{\'y}}}, \binits{M.}},
\bauthor{\bsnm{{B{\'a}rta}}, \binits{M.}}:
\byear{2008},
\batitle{{Fragmentation of the Current Sheet, Anomalous Resistivity, and
  Acceleration of Particles}}.
\bjtitle{\solphys}
\bvolume{247},
\bfpage{335}.
\doiurl{10.1007/s11207-008-9115-x}.
\adsurl{http://cdsads.u-strasbg.fr/abs/2008SoPh..247..335K}.
\end{barticle}
\endbibitem

\bibitem[\protect\citeauthoryear{{Karlick{\'y}} and
  {Yasnov}}{2015}]{2015A&A...581A.115K}
\begin{barticle}
\bauthor{\bsnm{{Karlick{\'y}}}, \binits{M.}},
\bauthor{\bsnm{{Yasnov}}, \binits{L.V.}}:
\byear{2015},
\batitle{{Determination of plasma parameters in solar zebra radio sources}}.
\bjtitle{\aap}
\bvolume{581},
\bfpage{A115}.
\doiurl{10.1051/0004-6361/201526785}.
\adsurl{http://cdsads.u-strasbg.fr/abs/2015A\%26A...581A.115K}.
\end{barticle}
\endbibitem

\bibitem[\protect\citeauthoryear{{Kuznetsov} and
  {Kontar}}{2015}]{2015SoPh..290...79K}
\begin{barticle}
\bauthor{\bsnm{{Kuznetsov}}, \binits{A.A.}},
\bauthor{\bsnm{{Kontar}}, \binits{E.P.}}:
\byear{2015},
\batitle{{Spatially Resolved Energetic Electron Properties for the 21 May 2004
  Flare from Radio Observations and 3D Simulations}}.
\bjtitle{\solphys}
\bvolume{290},
\bfpage{79}.
\doiurl{10.1007/s11207-014-0530-x}.
\adsurl{http://cdsads.u-strasbg.fr/abs/2015SoPh..290...79K}.
\end{barticle}
\endbibitem

\bibitem[\protect\citeauthoryear{{Kuznetsov} and
  {Tsap}}{2007}]{2007SoPh..241..127K}
\begin{barticle}
\bauthor{\bsnm{{Kuznetsov}}, \binits{A.A.}},
\bauthor{\bsnm{{Tsap}}, \binits{Y.T.}}:
\byear{2007},
\batitle{{Loss-cone instability and formation of zebra patterns in type IV
  solar radio bursts}}.
\bjtitle{\solphys}
\bvolume{241},
\bfpage{127}.
\doiurl{10.1007/s11207-006-0351-7}.
\adsurl{http://cdsads.u-strasbg.fr/abs/2007SoPh..241..127K}.
\end{barticle}
\endbibitem

\bibitem[\protect\citeauthoryear{{LaBelle}
  \textit{et~al.}}{2003}]{2003ApJ...593.1195L}
\begin{barticle}
\bauthor{\bsnm{{LaBelle}}, \binits{J.}},
\bauthor{\bsnm{{Treumann}}, \binits{R.A.}},
\bauthor{\bsnm{{Yoon}}, \binits{P.H.}},
\bauthor{\bsnm{{Karlick\'y}}, \binits{M.}}:
\byear{2003},
\batitle{{A model of zebra emission in solar type IV radio bursts}}.
\bjtitle{\apj}
\bvolume{593},
\bfpage{1195}.
\doiurl{10.1086/376732}.
\adsurl{http://cdsads.u-strasbg.fr/abs/2003ApJ...593.1195L}.
\end{barticle}
\endbibitem

\bibitem[\protect\citeauthoryear{Matsumoto and Omura}{1993}]{matsumoto.1993}
\begin{bbook}
\bauthor{\bsnm{Matsumoto}, \binits{H.}},
\bauthor{\bsnm{Omura}, \binits{Y.}}:
\byear{1993},
\bbtitle{{Computer space plasma physics: simulation techniques and software}},
\bpublisher{Terra Scientific},
\blocation{Tokyo},
\bfpage{305}.
\end{bbook}
\endbibitem

\bibitem[\protect\citeauthoryear{{Melrose}}{1985}]{1985srph.book..177M}
\begin{bbook}
\bauthor{\bsnm{{Melrose}}, \binits{D.B.}}:
\byear{1985},
In: \beditor{\bsnm{{McLean}}, \binits{D.J.}},
\beditor{\bsnm{{Labrum}}, \binits{N.R.}} (eds.)
\bbtitle{{Plasma emission mechanisms}},
\bpublisher{Cambridge University Press},
\blocation{Cambridge},
\bfpage{177}.
\adsurl{http://cdsads.u-strasbg.fr/abs/1985srph.book..177M}.
\end{bbook}
\endbibitem

\bibitem[\protect\citeauthoryear{{Peter} and
  {Bingert}}{2012}]{2012A&A...548A...1P}
\begin{barticle}
\bauthor{\bsnm{{Peter}}, \binits{H.}},
\bauthor{\bsnm{{Bingert}}, \binits{S.}}:
\byear{2012},
\batitle{{Constant cross section of loops in the solar corona}}.
\bjtitle{\aap}
\bvolume{548},
\bfpage{A1}.
\doiurl{10.1051/0004-6361/201219473}.
\adsurl{http://cdsads.u-strasbg.fr/abs/2012A\%26A...548A...1P}.
\end{barticle}
\endbibitem

\bibitem[\protect\citeauthoryear{{Peter}
  \textit{et~al.}}{2013}]{2013A&A...556A.104P}
\begin{barticle}
\bauthor{\bsnm{{Peter}}, \binits{H.}},
\bauthor{\bsnm{{Bingert}}, \binits{S.}},
\bauthor{\bsnm{{Klimchuk}}, \binits{J.A.}},
\bauthor{\bsnm{{de Forest}}, \binits{C.}},
\bauthor{\bsnm{{Cirtain}}, \binits{J.W.}},
\bauthor{\bsnm{{Golub}}, \binits{L.}},
\bauthor{\bsnm{{Winebarger}}, \binits{A.R.}},
\bauthor{\bsnm{{Kobayashi}}, \binits{K.}},
\bauthor{\bsnm{{Korreck}}, \binits{K.E.}}:
\byear{2013},
\batitle{{Structure of solar coronal loops: from miniature to large-scale}}.
\bjtitle{\aap}
\bvolume{556},
\bfpage{A104}.
\doiurl{10.1051/0004-6361/201321826}.
\adsurl{http://cdsads.u-strasbg.fr/abs/2013A\%26A...556A.104P}.
\end{barticle}
\endbibitem

\bibitem[\protect\citeauthoryear{{Priest}}{2014}]{2014masu.book.....P}
\begin{bbook}
\bauthor{\bsnm{{Priest}}, \binits{E.}}:
\byear{2014},
\bbtitle{{Magnetohydrodynamics of the Sun, \rm{Cambridge University Press,
  UK}}}.
\adsurl{http://cdsads.u-strasbg.fr/abs/2014masu.book.....P}.
\end{bbook}
\endbibitem

\bibitem[\protect\citeauthoryear{{Selhorst}, {Silva-V{\'a}lio}, and
  {Costa}}{2008}]{2008A&A...488.1079S}
\begin{barticle}
\bauthor{\bsnm{{Selhorst}}, \binits{C.L.}},
\bauthor{\bsnm{{Silva-V{\'a}lio}}, \binits{A.}},
\bauthor{\bsnm{{Costa}}, \binits{J.E.R.}}:
\byear{2008},
\batitle{{Solar atmospheric model over a highly polarized 17 GHz active
  region}}.
\bjtitle{\aap}
\bvolume{488},
\bfpage{1079}.
\doiurl{10.1051/0004-6361:20079217}.
\adsurl{http://cdsads.u-strasbg.fr/abs/2008A\%26A...488.1079S}.
\end{barticle}
\endbibitem

\bibitem[\protect\citeauthoryear{{Tan}}{2010}]{2010Ap&SS.325..251T}
\begin{barticle}
\bauthor{\bsnm{{Tan}}, \binits{B.}}:
\byear{2010},
\batitle{{A physical explanation of solar microwave zebra pattern with the
  current-carrying plasma loop model}}.
\bjtitle{Astrophys. Space Sci.}
\bvolume{325},
\bfpage{251}.
\doiurl{10.1007/s10509-009-0193-5}.
\adsurl{http://cdsads.u-strasbg.fr/abs/2010Ap\%26SS.325..251T}.
\end{barticle}
\endbibitem

\bibitem[\protect\citeauthoryear{{Tan}
  \textit{et~al.}}{2014}]{2014ApJ...790..151T}
\begin{barticle}
\bauthor{\bsnm{{Tan}}, \binits{B.}},
\bauthor{\bsnm{{Tan}}, \binits{C.}},
\bauthor{\bsnm{{Zhang}}, \binits{Y.}},
\bauthor{\bsnm{{Huang}}, \binits{J.}},
\bauthor{\bsnm{{M{\'e}sz{\'a}rosov{\'a}}}, \binits{H.}},
\bauthor{\bsnm{{Karlick{\'y}}}, \binits{M.}},
\bauthor{\bsnm{{Yan}}, \binits{Y.}}:
\byear{2014},
\batitle{{A very small and super strong zebra pattern burst at the beginning of
  a solar flare}}.
\bjtitle{\apj}
\bvolume{790},
\bfpage{151}.
\doiurl{10.1088/0004-637X/790/2/151}.
\adsurl{http://cdsads.u-strasbg.fr/abs/2014ApJ...790..151T}.
\end{barticle}
\endbibitem

\bibitem[\protect\citeauthoryear{{Thejappa}}{1991}]{1991SoPh..132..173T}
\begin{barticle}
\bauthor{\bsnm{{Thejappa}}, \binits{G.}}:
\byear{1991},
\batitle{{A self-consistent model for the storm radio emission from the sun}}.
\bjtitle{\solphys}
\bvolume{132},
\bfpage{173}.
\doiurl{10.1007/BF00159137}.
\adsurl{http://cdsads.u-strasbg.fr/abs/1991SoPh..132..173T}.
\end{barticle}
\endbibitem

\bibitem[\protect\citeauthoryear{{Watko} and
  {Klimchuk}}{2000}]{2000SoPh..193...77W}
\begin{barticle}
\bauthor{\bsnm{{Watko}}, \binits{J.A.}},
\bauthor{\bsnm{{Klimchuk}}, \binits{J.A.}}:
\byear{2000},
\batitle{{Width Variations along Coronal Loops Observed by TRACE}}.
\bjtitle{\solphys}
\bvolume{193},
\bfpage{77}.
\doiurl{10.1023/A:1005209528612}.
\adsurl{http://cdsads.u-strasbg.fr/abs/2000SoPh..193...77W}.
\end{barticle}
\endbibitem

\bibitem[\protect\citeauthoryear{{Wiegelmann}, {Thalmann}, and
  {Solanki}}{2014}]{2014A&ARv..22...78W}
\begin{barticle}
\bauthor{\bsnm{{Wiegelmann}}, \binits{T.}},
\bauthor{\bsnm{{Thalmann}}, \binits{J.K.}},
\bauthor{\bsnm{{Solanki}}, \binits{S.K.}}:
\byear{2014},
\batitle{{The magnetic field in the solar atmosphere}}.
\bjtitle{\aapr}
\bvolume{22},
\bfpage{78}.
\doiurl{10.1007/s00159-014-0078-7}.
\adsurl{http://cdsads.u-strasbg.fr/abs/2014A\%26ARv..22...78W}.
\end{barticle}
\endbibitem

\bibitem[\protect\citeauthoryear{{Yasnov} and
  {Karlick{\'y}}}{2015}]{2015SoPh..290.2001Y}
\begin{barticle}
\bauthor{\bsnm{{Yasnov}}, \binits{L.V.}},
\bauthor{\bsnm{{Karlick{\'y}}}, \binits{M.}}:
\byear{2015},
\batitle{{Regions of Generation and Optical Thicknesses of dm-Zebra Lines}}.
\bjtitle{\solphys}
\bvolume{290},
\bfpage{2001}.
\doiurl{10.1007/s11207-015-0721-0}.
\adsurl{http://cdsads.u-strasbg.fr/abs/2015SoPh..290.2001Y}.
\end{barticle}
\endbibitem

\bibitem[\protect\citeauthoryear{{Yasnov}, {Karlick{\'y}}, and
  {Stupishin}}{2016}]{2016SoPh..291.2037Y}
\begin{barticle}
\bauthor{\bsnm{{Yasnov}}, \binits{L.V.}},
\bauthor{\bsnm{{Karlick{\'y}}}, \binits{M.}},
\bauthor{\bsnm{{Stupishin}}, \binits{A.G.}}:
\byear{2016},
\batitle{{Physical Conditions in the Source Region of a Zebra Structure}}.
\bjtitle{\solphys}
\bvolume{291},
\bfpage{2037}.
\doiurl{10.1007/s11207-016-0952-8}.
\adsurl{http://cdsads.u-strasbg.fr/abs/2016SoPh..291.2037Y}.
\end{barticle}
\endbibitem

\bibitem[\protect\citeauthoryear{{Zaitsev} and
  {Stepanov}}{1983}]{1983SoPh...88..297Z}
\begin{barticle}
\bauthor{\bsnm{{Zaitsev}}, \binits{V.V.}},
\bauthor{\bsnm{{Stepanov}}, \binits{A.V.}}:
\byear{1983},
\batitle{{The plasma radiation of flare kernels}}.
\bjtitle{\solphys}
\bvolume{88},
\bfpage{297}.
\doiurl{10.1007/BF00196194}.
\adsurl{http://cdsads.u-strasbg.fr/abs/1983SoPh...88..297Z}.
\end{barticle}
\endbibitem

\bibitem[\protect\citeauthoryear{{Zheleznyakov}}{1997}]{1997riap.book.....Z}
\begin{bbook}
\bauthor{\bsnm{{Zheleznyakov}}, \binits{V.V.}}:
\byear{1997},
\bbtitle{{Radiation in astrophysical plasmas [in Russian]; Original Russian
  Title --- ``Izlucheniye v astrofizicheskoy plasme''}},
\bpublisher{Yanus-K},
\blocation{Moscow}.
\adsurl{http://cdsads.u-strasbg.fr/abs/1997riap.book.....Z}.
\end{bbook}
\endbibitem

\bibitem[\protect\citeauthoryear{{Zheleznyakov} and
  {Zlotnik}}{1975}]{1975SoPh...44..461Z}
\begin{barticle}
\bauthor{\bsnm{{Zheleznyakov}}, \binits{V.V.}},
\bauthor{\bsnm{{Zlotnik}}, \binits{E.Y.}}:
\byear{1975},
\batitle{{Cyclotron wave instability in the corona and origin of solar radio
  emission with fine structure. III. Origin of zebra-pattern.}}
\bjtitle{\solphys}
\bvolume{44},
\bfpage{461}.
\doiurl{10.1007/BF00153225}.
\adsurl{http://cdsads.u-strasbg.fr/abs/1975SoPh...44..461Z}.
\end{barticle}
\endbibitem

\bibitem[\protect\citeauthoryear{{Zlotnik}}{2013}]{2013SoPh..284..579Z}
\begin{barticle}
\bauthor{\bsnm{{Zlotnik}}, \binits{E.Y.}}:
\byear{2013},
\batitle{{Instability of electrons trapped by the coronal magnetic field and
  its evidence in the fine structure (zebra pattern) of solar radio spectra}}.
\bjtitle{\solphys}
\bvolume{284},
\bfpage{579}.
\doiurl{10.1007/s11207-012-0151-1}.
\adsurl{http://cdsads.u-strasbg.fr/abs/2013SoPh..284..579Z}.
\end{barticle}
\endbibitem

\end{thebibliography}

\end{article}
\end{document}